\documentclass[11pt,twoside]{article}
\usepackage{asp2004}
\usepackage{psfig}
\usepackage{epsf}
\usepackage{graphics}
\usepackage{lscape}
\newcommand{\gtsim}{\ {\raise-0.5ex\hbox{$\buildrel>\over\sim$}}\ }

\def\edcomment#1{\iffalse\marginpar{\raggedright\sl#1\/}\else\relax\fi}
\reversemarginpar

\begin{document}
\title{Constraining Dark Energy with the DEEP2 Redshift Survey}
\author{Marc Davis $^{1,2}$, Brian
F. Gerke$^2$, and Jeffrey A. Newman$^3$  \\
for the DEEP2 Team}
\affil{$^1$ Department of Astronomy, University of
California---Berkeley, USA \\
$^2$Department of Physics,  University of
California---Berkeley, USA \\
$^3$Lawrence Berkeley National Laboratory, USA}

\begin{abstract}

The DEEP2 survey has now completed half of its planned three-year
lifespan, and we have collected approximately $50\%$ of the data,
putting us exactly on schedule. The survey plan calls for
spectroscopic coverage by July of 2005 of $\sim$60,000 galaxies over
3.5 square degrees to a limiting magnitude of $R_{AB} = 24.1 $; the
great majority of these objects are at $0.7<z<1.4$.  We describe here
one method by which DEEP2 can set constraints on the equation of state
parameter of the Dark Energy, $w$.  By counting the number of
virialized groups and clusters we find in redshift space as a function
of their redshift and internal velocity dispersion, we probe both the
volume element and the growth of structure at $z\sim 1$, each of which
depends on $w$. Studies of early DEEP2 data indicate that the method
is likely to work, and preliminary indications are that a total of
$\sim 250$ small groups will be counted in the full survey, leading to
a constraint $\delta w \sim 0.1$ if combined with both the velocity
dispersion distribution of $z\sim 0$ clusters from the Sloan Digital
Sky Survey and an independent measurement of $\sigma_8$.  We also
provide a general description of the DEEP2 observations and target
selection, including the algorithm by which galaxies are placed on
slitmasks, to provide context for discussion of DEEP2 cluster samples.
\end{abstract}
\thispagestyle{plain}

\section{Introduction}
The DEEP2 (Deep Extragalactic Evolutionary Probe 2) survey is being
led by a consortium of astronomers at the University of California at
Berkeley and the University of California at Santa Cruz who are
pooling their Keck telescope time to gather spectra of $\sim$60,000
galaxies, most of which fall in the redshift range $0.7 < z < 1.4$.
As of this writing, we are pleased to announce that the survey is
halfway complete, and the list of scientific investigations currently
underway is longer than the list of collaborators ($\sim$ 2 dozen).
In this proceeding, we provide an update on prospects for 
constraining the Dark Energy equation of state parameter, $w$, by
counting groups of galaxies in DEEP2 as a function of their redshift
and velocity dispersion.  DEEP2 was designed to study galaxies over
several, independent, comparatively wide fields; this will provide large
enough samples and small enough cosmic variance to measure $w$ using
large-scale structure in the DEEP2 fields.

We will begin our discussion by briefly summarizing the observational
program of the DEEP2 survey.  The survey will be described in more
detail in an upcoming paper by Faber et al. (2004).  We then discuss
the theoretical basis for measuring $w$ using counts of galaxy groups.
To provide a preview of the measurement we will make with the final
DEEP2 catalog, we identify and count groups in the current DEEP2 data
and compare these measurements to expectations from realistic DEEP2
mock catalogs produced from N-body simulations.  We finish by
computing the cosmological constraints that are expected from such a
measurement.

\section{Details of the Observations}

The University of California has guaranteed DEEP2 80 nights of dark
observing time at the Keck Observatory over 3 years.  The survey is
made possible by DEIMOS, a new multi-object spectrograph on the Keck 2
telescope (Faber et al. 2002).  DEIMOS' large field of view (16'
$\times$ 5'), large detector mosaic (8k $\times$ 8k pixels, allowing
broad 
spectral coverage at high resolution), and active-feedback Flexure
Compensation System (which allows the effects of fringing to be
removed, greatly improving sky subtraction) are all key to our
survey strategy and its success.  DEIMOS was commissioned in June
2002, and the DEEP2
survey began immediately thereafter, in July.  This review is written after 3
semesters of observing, out of a total of 6; current plans call for
observing to be completed in July, 2005.  At the halfway
point, we have observed 245 of 480 DEIMOS slitmasks, and collected
spectra for $\sim$30,000 galaxies.  Needless to say, our small team is
barely keeping up with the huge influx of data!

\subsection{The DEEP2 Fields}

The Survey is designed to observe during the months of April through
October at the Keck observatory on Mauna Kea, Hawaii. There are 4
fields spaced by $\sim 60$ degrees, allowing us always to move from
one field to the next while the fields are at small zenith angle.  The
3.5 square degrees of field coverage is divided amongst our four
fields spread across the sky as shown in Table~\ref{tab:fields}. For
the first field, we target galaxies with $R_{AB} < 24.1$ in an 120' by
16' region of sky.  The long axis of this field, known as the Extended
Groth Strip (EGS), is oriented along a line of ecliptic latitude,
allowing extremely efficient mid-IR imaging with the \emph{Spitzer
Space Telescope}.  It is the subject of some of the deepest wide-area
($120' \times 10'$, embedded within the region observed by DEEP2)
\emph{Spitzer} imaging on the sky, currently being conducted by the
MIPS and IRAC instrument teams (Fazio et al., 2004; 
Papovich et al., 2004). This field is also the focus of a
large \emph{Hubble Space Telescope} Advanced Camera for Surveys
campaign now underway, covering a $60' \times 10'$ region in the
middle of the \emph{Spitzer} coverage in both $V$ and $I$, along with deep
\emph{GALEX}, \emph{Chandra}, and \emph{XMM} imaging.


\begin{table}[!t]
\smallskip
\begin{center}
{\small
\begin{tabular}{ccccc}
\tableline
\noalign{\smallskip}
Field & RA  & $\delta$ & area &  fraction  \\
Number &    &        & (deg$^2$) & observed  \\      
\noalign{\smallskip}
\tableline
\noalign{\smallskip}
1 & 14$^h$ 17$^m$  & +52 30      &       0.5    &   0.25  \\ 
2 & 16$^h$ 52$^m$    & +34 55        &       1.0    &       0.35  \\
3 &   23$^h$ 30$^m$ & +0        &       1.0    &       0.70 \\
4 &    2$^h$ 30$^m$ & +0        &         1.0  &      0.50  \\
\noalign{\smallskip
}
\tableline

\end{tabular}
}
\caption{Summary of the DEEP2 fields. Shown are the
locations, areas on the sky, and currently observed fraction for each of our four
fields.}
\label{tab:fields}

\end{center}
\end{table}

The remaining fields of DEEP2 use deep BRI imaging obtained with the
CFH12K camera on the Canada France Hawaii Telescope (Coil et al. 2004)
to pre-select galaxies with $z>0.7$ by means of a simple color cut
(Newman et al. 2004).  This pre-selection allows us to observe
galaxies at high redshift efficiently; based on studies using data
from the Extended
Groth Strip (where no color cut is applied), more than 60\% of
galaxies with $R_{AB}<24.1$ are at 
$z<0.75$, but only 13\% of galaxies that pass the color cut are; while
only 3\% of objects rejected are at $z>0.75$.  Each of these fields
comprises three imaging pointings of CFHT, covering a total area of
120' by 30' and oriented with the long direction East-West.

\subsection{The DEEP2 Sample}

There are approximately $10^5$ galaxies with $R_{AB} < 24.1$ and $z >
0.7$ in the 3.5 sq. degrees surveyed; because of slitmask design
constraints and the finite amount of observing time available, we are
able to obtain spectra for only $\sim$60,000 of these potential
targets.  Of the observed galaxies, we will successfully obtain
redshifts for $\sim$50,000, and resolved kinematics (e.g. rotation
curves) for $\gtsim$5000.  Based on follow-up studies, it appears that
the majority of DEEP2 redshift failures are objects at $z>1.4$, which
lack features in our spectroscopic window (Steidel,
private communication 2003).  Crudely, one can think of the DEEP2
sample as 
having twice the number of objects and 3 times the volume as the Los
Campanas Redshift Survey (Shectman et al. 1996), but
at $z\sim 1$ rather than $z\sim 0$.  The total volume
to be surveyed ($\sim6 \times 10^6 h^{-3} \mathrm{Mpc}^3$ comoving,
assuming a standard $\Lambda$CDM cosmology) should not contain many
extremely rich clusters, but is sufficient for counting less extreme
objects like galaxies and groups of galaxies.  Having four independent
fields both limits the impact of and allows us to measure the effects
of cosmic variance.

In studying surveys which cover a broad redshift range such as DEEP2,
it is important to understand that such a sample will inevitably
select different sorts of galaxies at different redshifts.  For
instance, the $R$ band in which the DEEP2 sample is magnitude-limited
corresponds to restframe $\sim 4000 \AA$ at $z\sim 0.7$, but to $\sim
2800 \AA$ at $z\sim 1.4$.  Thus our wide redshift range means that
galaxies are not selected based on the same rest-frame properties at
different redshifts; one must be mindful of how homogeneous any sample
of galaxies covering a broad redshift range actually is.

For instance, galaxies with ``early-type'' spectra (i.e., spectra
which resemble those of low-redshift elliptical and S0 galaxies) are
redder than late-type galaxies that are forming stars; so for a given
rest-frame $B$ or $R$ absolute magnitude and redshift, an early-type
galaxy will be fainter than a late-type galaxy in the observed $R$
band.  This effect becomes stronger as one moves to higher redshift
and the rest-frame band corresponding to observed $R$ becomes bluer.
It is important to note that this effect cannot be removed simply by
selecting objects in redder wavebands.
Even surveys which target galaxies for spectroscopy based on infrared
magnitudes (e.g. $K$) are likely not to measure redshifts for objects
that are faint in the optical, so they too will tend to lose
intrinsically red objects from their sample before bluer ones due to
redshift failures, while simultaneously only the brightest
intrinsically blue objects will fulfill the sample's selection
criteria.  As a practical matter, for DEEP2 samples these effects mean
that our sample probes deeper into the luminosity function for blue
galaxies (compared to the $L_*$ for blue galaxies) than it does for
red galaxies (compared to the $L_*$ for red galaxies; for $z \gtsim
1.1$, early-type galaxies are all but absent from the DEEP2 sample.

Although we define which galaxies we wish to observe by means of the
magnitude limit and color cut, it is impossible to place all of them
on DEIMOS slitmasks for observation.  No two slitlets can have
overlapping spectra (as otherwise the skylines from one hopelessly
contaminate the other), and it is all but impossible to achieve
effective sky subtraction for slits less than 3 arcsec long; these two
requirements limit the number of slitlets we can place on a mask.
These constraints imply that two objects separated by less than
$\sim$3 arcsec in the direction parallel to the long axis of a mask
cannot both be observed simultaneously, regardless of their location
in the short (4') direction.  

To mitigate this effect, nearly every position on the sky in the DEEP2
survey 
is covered by at least 2 masks, giving objects multiple opportunities
to be selected.  We also assign each object a weight $W$ between 0 and 1
according to its probability of being a galaxy as opposed to a star
(cf. Coil et al. 2004), its $R$ magnitude (if fainter objects are not
deweighted somewhat, the sample will tend to pile up at the magnitude
limit of the survey), and its consistency with the color cut (outside
the nominal cut, the $W$ falls off as a Gaussian with
$\sigma=0.05$; this provides a 'prewhitening' of the color selection
at a level greater than our estimated systematics in $B-R$/$R-I$
space).  This weight is used for randomly selecting
amongst objects that would have overlapping spectra, as discussed
below, allowing us to
focus on objects with the desired characteristics while still sampling
a wider parameter space.

For fields other than the Extended Groth Strip, objects are al\-lo\-cat\-ed
among these multiple slitmasks through a two-stage procedure (in the
EGS, each point on the sky is covered by 4 masks with two different
orientations; we thus use a somewhat more complicated algorithm than
that described here, though it is largely analogous).  The first
stage, or 'pass', places slitlets only in a central region unique to
each mask -- 2 arcmin wide on average, though this varies somewhat due
to the adaptive tiling of masks (q.v. below).  This central region is
favored because galaxies at one end of the mask or the other in the
short axis will have different wavelength coverage on the detector (as
this axis corresponds to the spectral direction in DEIMOS). 
The wavelength shift is $\sim 100$ \AA/arcmin, so that this is not a
very strong effect, but nonetheless one of which we are mindful.  
The central target selection region for each slitmask is not actually
rectangular on the sky, but instead is bowed parallel to a line of
constant central wavelength on the DEIMOS detector (which is not
straight on the sky due to optical distortions).

In the first pass, we initially trim the list of targets by generating
a random number between 0 and 1 and only retaining those for which the
random number is less than their weight, so low-$W$ objects
(e.g. those which are likely to be stars) are only rarely placed on
slits in the central portion of a mask.  We then search this trimmed
list for cases where we can place two objects (separated by less than
3 arcsec in the long direction of the mask) on a single slit which has
PA relative to the mask of less than $30^\circ$.  Slits are allocated
to all such cases, boosting our ability to study close pairs of
galaxies.  We then generate a random priority, $P_1$, for each of the
remaining objects in the trimmed list.  Each one that can be observed
without precluding any other is always selected for observation, while
in cases of conflict the object with greatest $P_1$ is taken.  This
procedure (initial selection according to $W$, random selection
amongst the surviving targets) yields a sample where the probability
of selection is directly proportional to an object's weight, making
the best possible use of the central region of each mask.

In the second pass, objects are selected over the full 16' $\times$ 4'
area covered by the mask, rather than just the central region.
Objects are given a priority in the second pass, $P_2$, equal to their
weight, $W$, divided by a random number between 0 and 1.  We then
assign slits to objects in decreasing order of $P_2$, until every
object either has been selected or conflicts with a slit on that mask.
Since any target object that can be placed on a mask without causing a
conflict is put on that mask, in the second pass low-$W$ objects
will often be selected for observation, since doing so is cost-free.
The second pass proceeds from west to east amongst the masks, in
order.

The second pass again ranks galaxies in priority using their weights
and random numbers, but now over the full 16' $\times$
4' area covered by the mask, rather than just the central region.
Slits are then allocated to galaxies in descending order of priority,
so long as they do not conflict with any slit that has already been
added to the mask.  In the end, any target object that can be placed
on a mask without causing a conflict is put on that mask, so that in
the second pass low-weight objects will frequently be selected for
observation, since they are cost-free.  The second pass proceeds from
west to east 
amongst the masks, in order.

Cosmic variance in the number of targets within a 16' $\times$ 2'
region is high; if masks are evenly spaced, some will have many fewer
slits than others.  To avoid this, we iteratively adjust the
positions (and, correspondingly, the widths of the central region) of
each mask such that the number of targets selected in the first pass is
constant mask-to-mask.  This enables us to obtain spectra for a
uniform fraction of galaxies over the survey region.

In the end, we select $> 60\% $ of the eligible
galaxies for spectroscopy.  In using such a sample to study
large-scale structure, it is important to realize that the selection
of targets for spectroscopy causes the two-point autocorrelation
function $\xi(r)$ to be distorted (Coil, Davis, \& Szapudi 2001).  
These authors show that the effect on the measured
$\xi(r_p,\pi)$ is modest, with the principal impact being a reduction
in the elongation of contours of $\xi$ at small separation due to
'fingers of god', corresponding to a reduction in velocity dispersion
by $\sim 20$\%.  Figure 1 shows actual mask designs for a portion
of one DEEP2 field.

At the end of this process, each slitmask contains $\sim$110-150
slitlets, typically $\sim$7 arcsec long but spanning the range $\sim
3-15+$ arcsec, over a 16' by 4' region.  The long axis of an
individual slitlet is oriented along the major axis of each galaxy
measured from the CFHT photometry, so long as that axis is within $\pm
30$ degrees of the long axis of the slitmask and the galaxy is
measurably ellipsoidal.  Each mask is observed for at least 3
20-minute exposures with the 1200 l/mm grating on DEIMOS, centered at
7800 \AA.  This setting provides comparatively high-resolution ($R
\gtsim 5000$) spectroscopy over a 2600 \AA ~ window.  Observing at high
resolution has two major advantages: it allows us to split the [OII]
3727 \AA ~ doublet feature, providing a secure redshift for
line-emitting objects at $0.7<z<1.4$ even if that is the only feature
available; and it effectively provides OH suppression, since only
$\sim 20\%$ of all pixels are influenced by skylines in this setting,
yielding large domains of skyline-free spectrum (our sky subtraction is
effective on the skylines as well; if we compare the redshift
distribution of DEEP2 sources to the sky spectrum, matching
wavelength to redshift by assuming that the $z$ was measured from
[OII], we find a correlation coefficient of only -0.1) 
Under optimal conditions, we can
observe 8 masks per night, which will yield spectra of $\sim$1000-1200
galaxies.

\section{Using the DEEP2 Survey to Study Dark Energy} 

DEEP2 can set constraints on the dark energy through two variants on a
classical cosmological test, the 'dN/dz test'.  This test can
constrain cosmological parameters like the mass and dark energy
density parameters $\Omega_m$ and $\Omega_X$, or the dark energy
equation of state parameter $w$, by measuring the apparent abundance
per unit redshift and solid angle, $dN/dz$, of a class of objects.
The abundance depends on fundamental cosmological parameters and comoving
number density via the relation
\begin{equation}
dN/dz = n(z) dV/dz \propto n(z) r(z)^2 / E(z) dz,
\end{equation}
where $z$ is the redshift of interest; $n(z)$ is the comoving number
density of this class of object at that redshift; $dV/dz$ is the
amount of comoving volume per unit redshift and solid angle; $E(z)$ is
the familiar Hubble ratio, given by
\begin{equation}
E(z) \equiv H(z)/H_0 =
(\Omega_m(1+z)^3+\Omega_X(1+z)^{3(1+w)})^{1/2}
\end{equation}
at late times (and assuming that $w$ is independent of $z$); and
$r(z)$ is the comoving distance to redshift $z$, 
\begin{equation}
r(z) = (c/H_0) \int{dz/E(z)}.
\end{equation}
The relationship between number counts and
cosmology formed the basis of some of the earliest attempts to
determine the geometry of the Universe (Hubble 1926).

In general, this test has been applied in one of two limits: where
$n(z)$ is presumed to be known (e.g. Loh \& Spillar 1986), or where
$n(z)$ is much more sensitive to cosmology than the volume element
$dV/dz$ (e.g. studies of X-ray cluster abundances,  [Borgani et al. 2004;
Bahcall et al. 1997]).  The DEEP2 Redshift Survey has the potential to
provide 
cosmological constraints spanning both of these limits.  Newman \&
Davis (2000) showed that the comoving abundance of dark matter halos
with circular velocity $\sim 200$ km/sec -- those which presumably
host typical $L_*$ galaxies -- normalized to the abundance observed at
$z\sim 0$ is almost entirely independent of cosmological parameters.
However, the application of this method is limited by our ability to
predict the general effects of baryonic infall (Newman \& Davis 2002)
and the lack of a matching sample at $z \sim 0$ for normalization
(local, fiber-based surveys such as SDSS only measure the linewidth of
gas at the centers of galaxies, which may not reflect the kinematics
of the dark matter halo); we therefore are concentrating on other
methods in the short term.

In contrast to typical galaxies, $n(M,z)$, the comoving abundance of
rich galaxy clusters as a function of their mass, $M$, and redshift,
$z$, is exponentially sensitive to the growth rate of large-scale
structure, which depends on cosmological parameters including $w$
(Linder \& Jenkins, 2003).  Exploiting this fact, the $dN/dz$ test has been
applied by using X-ray properties or optical richnesses of clusters,
and should be possible with other methods (e.g. blind
Sunyaev-Zel'dovich effect surveys) in the future. 

Apart from the difficulty of finding them in X-ray or optical
photometric surveys, there is no fundamental reason why poorer groups
cannot be incorporated into such analyses; because their comoving
abundance is less sensitive to structure formation than more massive
objects, they provide constraints in different directions in the
$\Omega_m--w$ plane than more massive clusters, so a simultaneous
analysis will yield more information than a study of rich clusters
alone.  However, the only currently effective method for finding less
massive groups with few luminous members is to detect them in redshift
space, as their X-ray brightness is low (especially at higher
redshift) and their contrast against background galaxies in surface
density is poor.

As a free byproduct, the group line-of-sight velocity dispersion,
$\sigma$, which roughly obeys the relation $\sigma \propto M^{1/3}$,
is determined in the process of finding groups in redshift space.  The
velocity dispersion distribution of dark matter halos can be predicted
either from analytic methods like Press-Schechter or N-body
simulations, much like the mass distribution; in hydrodynamic
simulations, it is much more strongly correlated with cluster mass
than gas diagnostics like X-ray temperature (Evrard 2004).  
Understanding the details of the relationship between
what is measured and the true halo velocity dispersion still poses
some challenges---\emph{e.g.}, the nature and extent of galaxy
``velocity bias'' remains uncertain.  However, these topics are the
subject of a variety of simulation efforts, and we have every
expectation that a resolution should be reached in the next few years;
observational results will stand unchanged, but their interpretation
could improve over time.

As suggested by Newman et al. (2002), the DEEP2 Redshift Survey is now
providing data on groups to $z\sim1.4$ that can be used for this
test.  We have implemented a sophisticated cluster-finding algorithm
(Marinoni et al.\ 2002, Gerke et al.\ 2004) which finds groups adaptively
using the locations of galaxies in redshift space. This removes most
of the background contamination problems of photometric methods, but
not all (since galaxies cluster with each other and velocity
dispersions can correspond to $\sim$20 Mpc in length, interloper
contamination is inevitable).  Furthermore, as described in $\S 2.2$,
we cannot get spectroscopy for every object, especially in the densest
regions; e.g., sometimes we will lose 2 members out of a 3-member group by
chance, causing the group not to be identified. 
However, extensive tests with mock catalogs (Yan et al. 2004) have
shown that we recover the \emph{velocity distribution} of clusters for
$\sigma > 400$ km/sec to well within the expected cosmic variance in
group counts, and that we find groups equally effectively across the
DEEP2 redshift range and over a wide span in velocity dispersion,
yielding a nearly unbiased sample.

Some results from our search for galaxy groups can be seen in Figure 2, 
which shows the measured group velocity distribution from 1/12th
of the total planned DEEP2 dataset.  Random errors for the full DEEP2
sample will be only $\sim$1/3 as large.  Also plotted are the average
actual $n(\sigma)$ distribution for groups in twelve mock DEEP2
samples
from Yan et al. (2004); and the RMS variation of those mocks about
that line, due to Poisson statistics and cosmic variance (shaded
region).  As can be seen in Figure 3, the typical group found in
DEEP2 has only a few members luminous enough to be targeted for
spectroscopy ($\sim L_*$ or brighter at $z\sim 1$), so the contrast
between these groups and the field is extremely poor when there is no
redshift information available.

Some examples of the difficulties in identifying poor groups and their
members are shown in Figure 4, which depicts two real groups and
two groups from a mock catalog plotted on the sky, as well as in
redshift space.  It is all but impossible to tell the difference; each
real group has a doppelganger in the group catalog from simulations.
The fingers of god are obvious and are much larger than our redshift
precision, $\sim 30$ km/sec. The estimation of group velocity
dispersion is clearly straight-forward; in the long term, we will be
able to compare these to X-ray temperatures (many weak groups can be
stacked to yield strong signal-to-noise), Sunyaev-Zel'dovich
decrements, and lensing mass (at the low-redshift end), especially in the Extended Groth Strip where wide arrays of complementary data are avalilable.

We expect to find $\sim$250 groups and clusters in the full DEEP2
sample. As described by Newman et al. (2002), under the assumptions that
$\sigma_8$ is known independently of cluster studies (e.g. by using
galaxy-galaxy lensing to turn the clustering of galaxies into a
clustering of mass, and hence $\sigma_8$), that the velocity
dispersion distribution of groups at $z \sim 0$ has been measured by
the Sloan Digital Sky Survey, and that the geometry of the
Universe is flat, it is possible to constrain the dark energy
equation of state parameter with a precision of $\delta w/w \sim 0.1$.
We have recently started refining that analysis to prepare for
applying these methods to the DEEP2 data; amongst other things, we
find that the precision of this measurement is be degraded only
slightly when completely covariant systematic errors are incorporated
into the analysis of Newman et al.
We eagerly await the first results from this procedure, which we will
tackle as soon as we finish observing and determining redshifts for
DEEP2 galaxies, in mid-to-late 2005. 

\acknowledgments{We thank our collaborators Sandy Faber, David Koo,
and Puragra Guhathakurta and  the  
hardworking teams at Berkeley and Santa Cruz.
This project was supported by the NSF grant AST-0071048.  BFG
acknowledges support from an NSF Graduate Research Fellowship, and  
JAN acknowledges support by NASA through Hubble Fellowship grant
HST-HF-01132.01 awarded by the Space Telescope Science Institute, which
is operated by AURA Inc. under NASA contract NAS-5-26555. 
}

\begin{figure}[!t]
\centering 
\plotfiddle{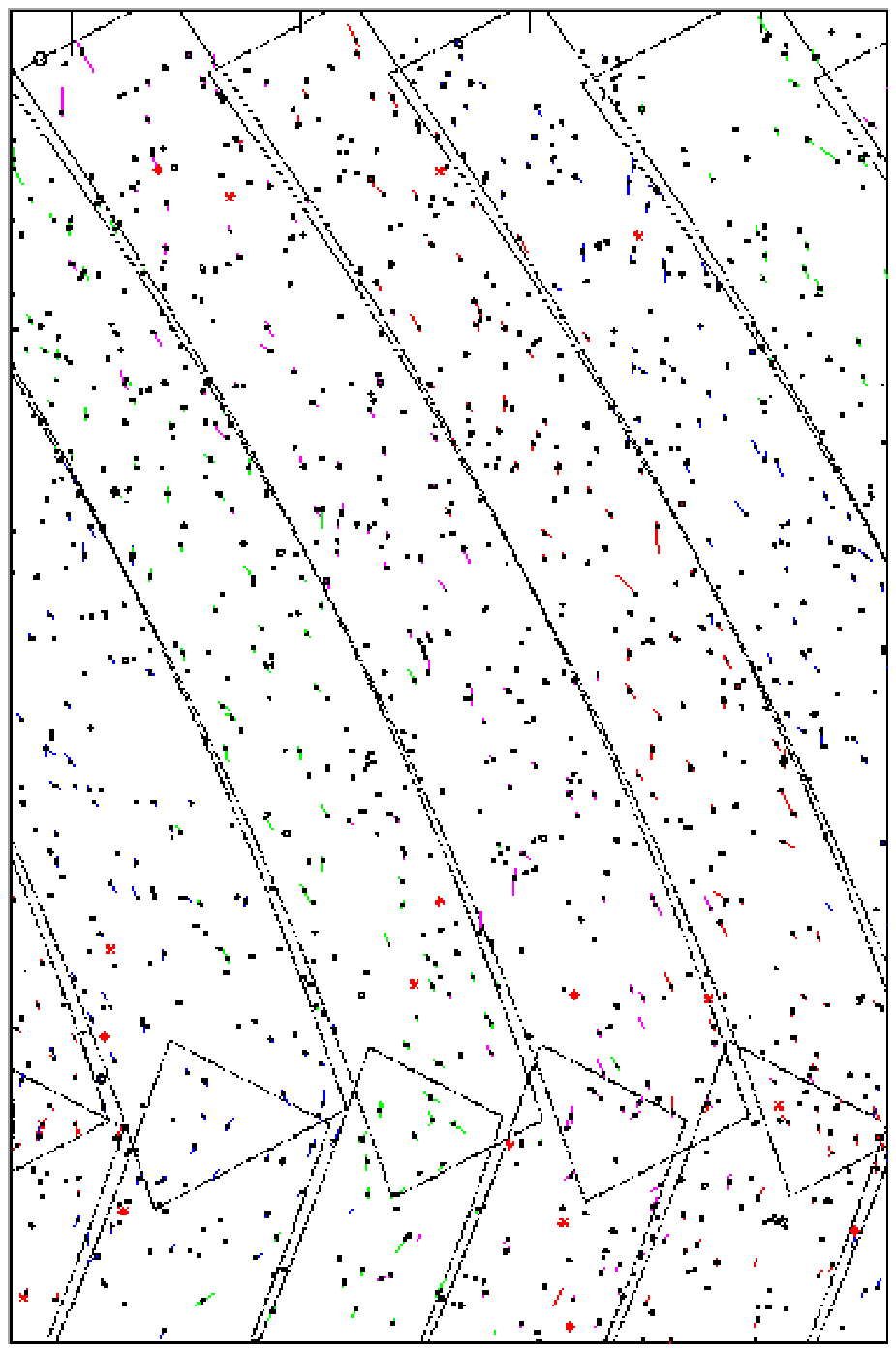}{6in}{0}{100}{100}{-250}{-250}
\caption{DEEP2 targets, slitlets, and masks in an $\sim 18' \times
10'$ region drawn 
from a 42' $\times$ 28' CFHT pointing.  
Each pointing is tiled
with two slightly-overlapping rows of masks, with the masks tilted to
minimize the effects of atmospheric dispersion at the time of
observation.  Target selection for each row is done separately,
allowing some objects to be observed twice for data quality testing.
The dotted lines roughly indicate the central, ``pass one'' portion
of each mask; ellipses show eligible DEEP2 targets.  Each small
rectangle plotted shows a DEEP2 slitlet to scale; its color indicates
which mask the slit was placed on (note that most slits of the same
color fall in the central portion of the mask corresponding to that
color, but not all).
}
\label{fig:maskmaking}
\end{figure}

\begin{figure}[!t]
\centering
\plotone{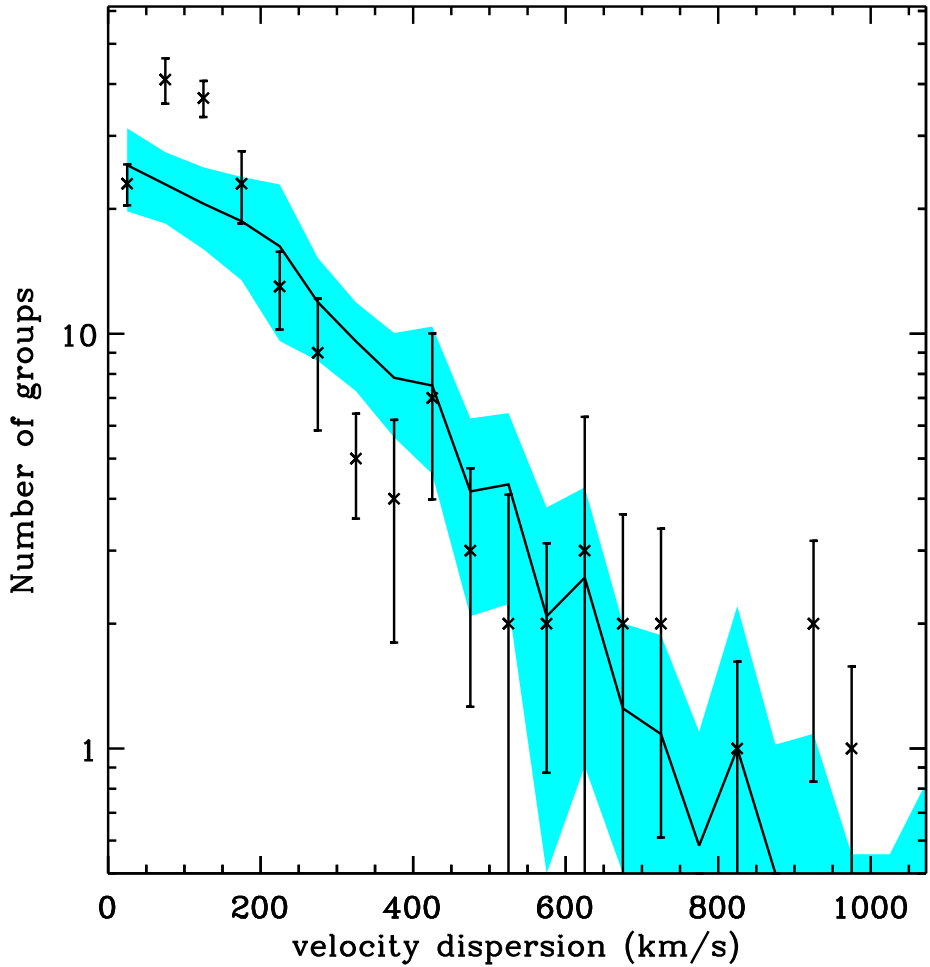}
\caption{Comparison of the velocity function $n(\sigma)$ measured in
a single DEEP2 pointing to that predicted by mock catalogs.  The data
points are the measured velocity function
of groups found in the most uniformly observed DEEP2 pointing
(Figures~\ref{fig:clust_sky}),
in bins of $50$ km/s.  Error bars are estimated from applying the VDM
group finder to twelve independent mock DEEP2 pointings, measuring
$n(\sigma)$, and taking the standard deviation of the residuals
$n_{\mathrm{found}}-n_{\mathrm{true}}$ . The solid line is the average
velocity function
from the mock catalogs, $\langle n_{\mathrm{true}}(\sigma)\rangle$,
in bins of $50$ km/s, and the shaded
region indicates the cosmic variance (standard deviation) in each
bin.  The measured velocity function is consistent with the prediction
in the regime ($\sigma > 400$ km/s) where an accurate measurement is
expected based upon the mock catalogs.}
\label{fig:dist_comp}
\end{figure}

\begin{figure}[!t]
\centering
\plotone{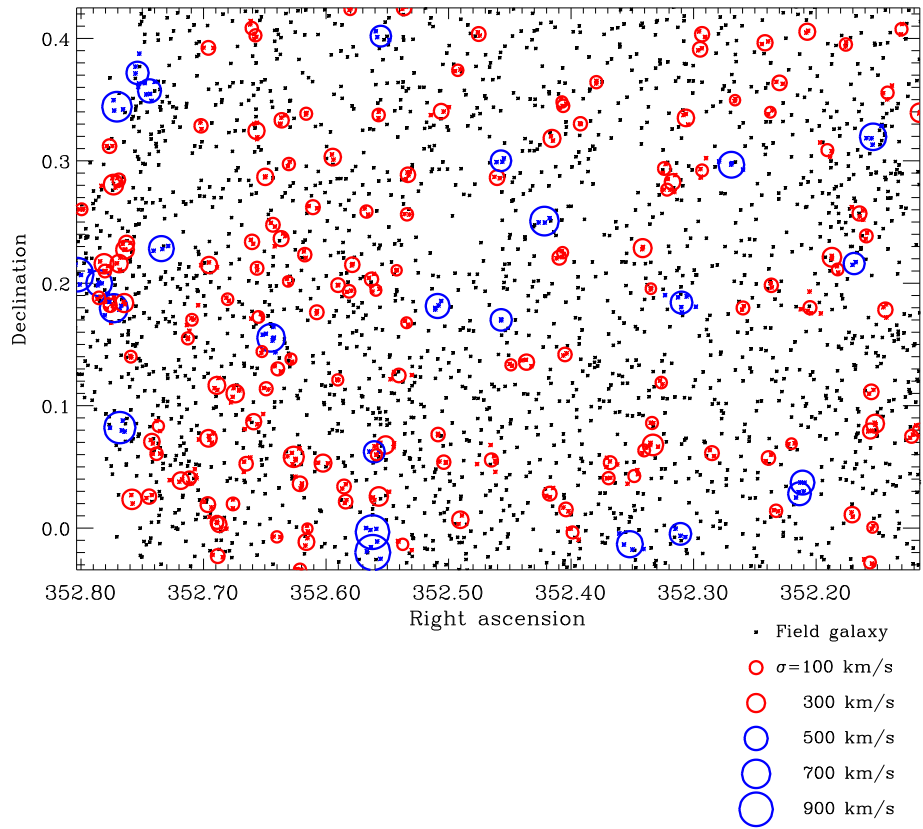}
\caption{Groups in the most uniformly observed DEEP2 field, as seen on
the sky.  Colored circles indicate the position of the groups, with
circle radius proportional to group velocity dispersion.  Red circles
denote groups with velocity dispersion $\sigma < 400$ km/s, and blue
circles show $\sigma \ge 400$ km/s groups.  Similarly, colored dots
indicate group member galaxies.  Isolated galaxies are shown by black
dots.}
\label{fig:clust_sky} 
\end{figure}

\begin{figure}[!t]
\centering
\plotfiddle{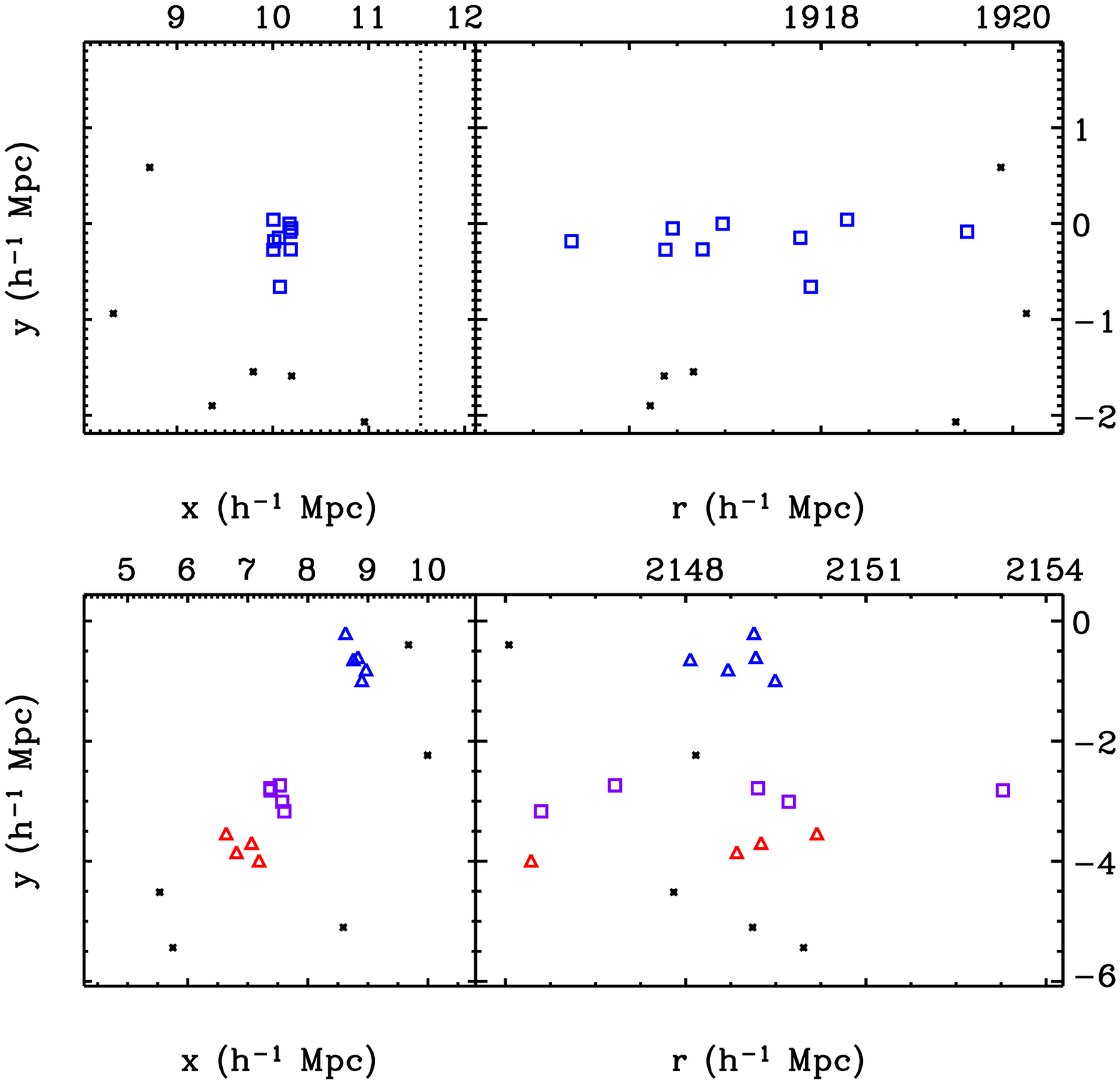}{2in}{0}{35}{35}{-100}{-20}
\plotfiddle{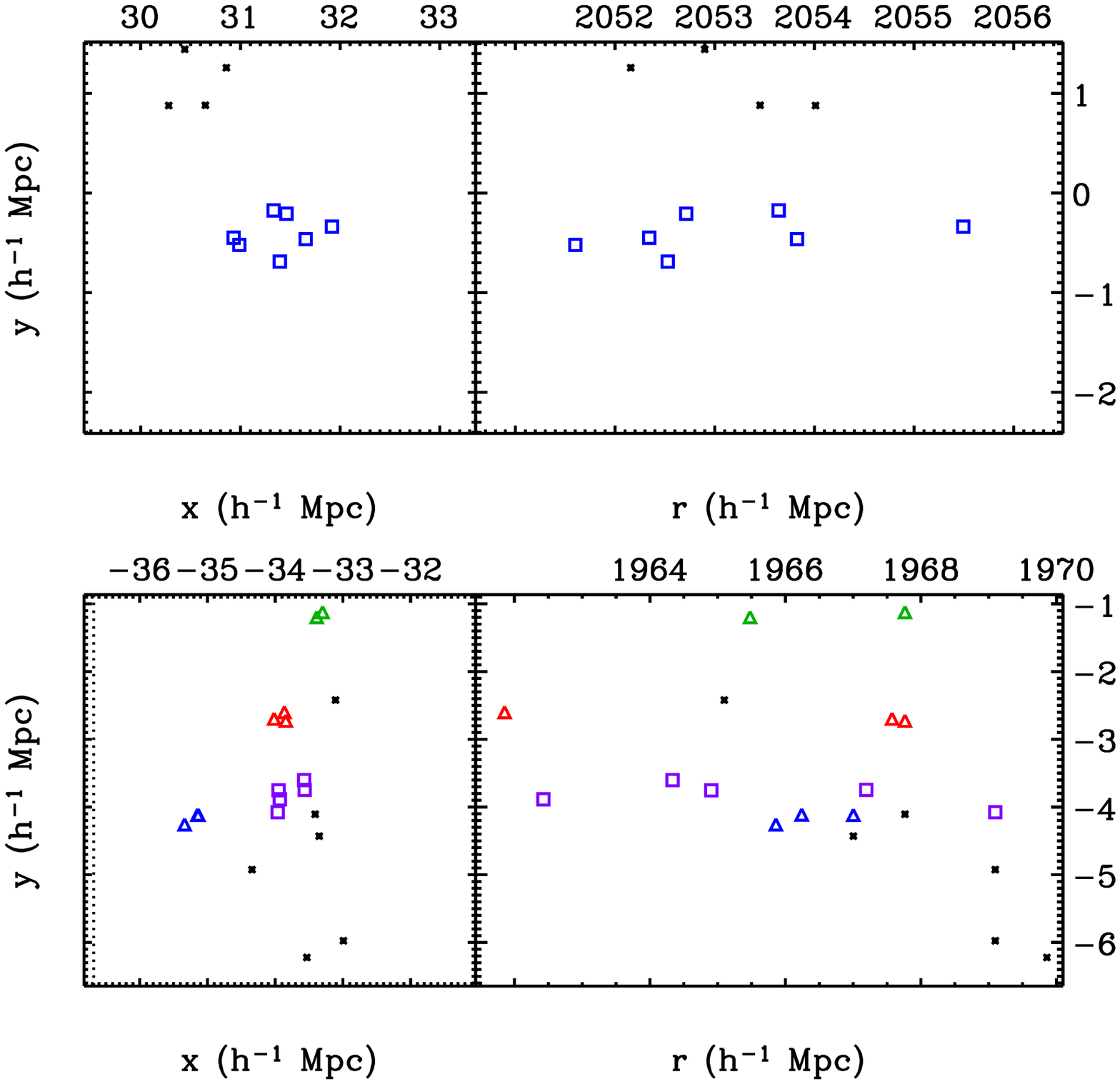}{2in}{0}{35}{35}{-100}{-60}
\caption{A selection of DEEP2 groups, shown as seen on the sky and
projected in redshift space.  Colored squares denote the galaxies in
the group being shown; colored triangles show galaxies in nearby
groups, and black points show nearby isolated galaxies.  Dashed lines
indicate the edge of the survey field.  Two of the
groups shown here come from real DEEP2 data, and 
two come from mock catalogs.  Distinguishing between them is left as
an exercise for the reader.}
\label{fig:group_plot}
\end{figure}


\end{document}